\documentclass[a4paper,11pt]{article}
\usepackage{pos}
\usepackage{natbib}
\title{Latest D0 results on exotic hadrons produced in $p\bar{p}$ collisions}
\author*[a]{Alexey Drutskoy}
\affiliation[a]{P. N. Lebedev Physical Institute of the Russian Academy of Sciences,\\
  53 Leninskiy Prospect, Moscow, Russia}

\emailAdd{drutskoy@lebedev.ru}
\abstract{Prompt and nonprompt productions of exotic multiquark states
are studied using the $\sim$10.4~fb$^{-1}$ data sample collected
by the D0 experiment in Tevatron $p\bar{p}$ collisions at \mbox{$\sqrt{s}$ =} 1.96 TeV.
The recent D0 results on the prompt and nonprompt production of the
$X(3872)$ and $Z_c^+(3900)$ states and the $P_c$ pentaquarks at the 4450~MeV region
are reported.
Signals corresponding to these states are found in the nonprompt production,
whereas only the $X(3872)$ state is seen in the prompt production.
The ratio of prompt to nonprompt $X(3872)$ production 
is about three times larger in the D0 measurement than that obtained by 
the ATLAS experiment at 8~TeV.
Theoretically, the production, formation, coalescence, and disassociation processes
are expected to be quite different
for conventional mesons with a spatial size of (0.4--0.8) fm,
compact multiquark states such as tetraquarks with a size of a few fm,
and spatially extended molecular states with a size of \mbox{(4--10) fm}.
They can be differently affected in prompt hadron-hadron collisions
where there are many additional particles emitted from the interaction point.
Consequently, the prompt to nonprompt production ratio
of spatially extended exotic states
can be suppressed at LHC comparing with the Tevatron conditions,
because of large difference in the hadron-hadron collisions particle multiplicity.
The prompt production studies provide an opportunity to better understand
the nature of exotic states.}

\FullConference{%
  40th International Conference on High Energy physics - ICHEP2020\\
  July 28 - August 6, 2020\\
  Prague, Czech Republic (virtual meeting)
}

\begin{document}
\maketitle

\section{Introduction}

Since the discovery of the state $X(3872)$ (also named $\chi_{c1}(3872)$)
in 2003 by the Belle
collaboration~\cite{bel}, many new exotic states have been observed.
Whereas the conventional states consist of a quark-antiquark
pair or three quarks, exotic states include a quark-antiquark pair
in addition to the conventional configurations.
Theoretically, the exotic states can be described as compact tetraquark
(or pentaquark) states or spatially extended molecular configurations.
The spatial dimensions of these states are very different and usually
estimated to be about (0.4--0.8)~fm for conventional states,
(1--4)~fm for compact tetraquarks, and (4--10)~fm for molecular states.
This can have a significant impact on production processes of these states.

\begin{figure}[h!]
\begin{center}
\includegraphics[width=.35\textwidth]{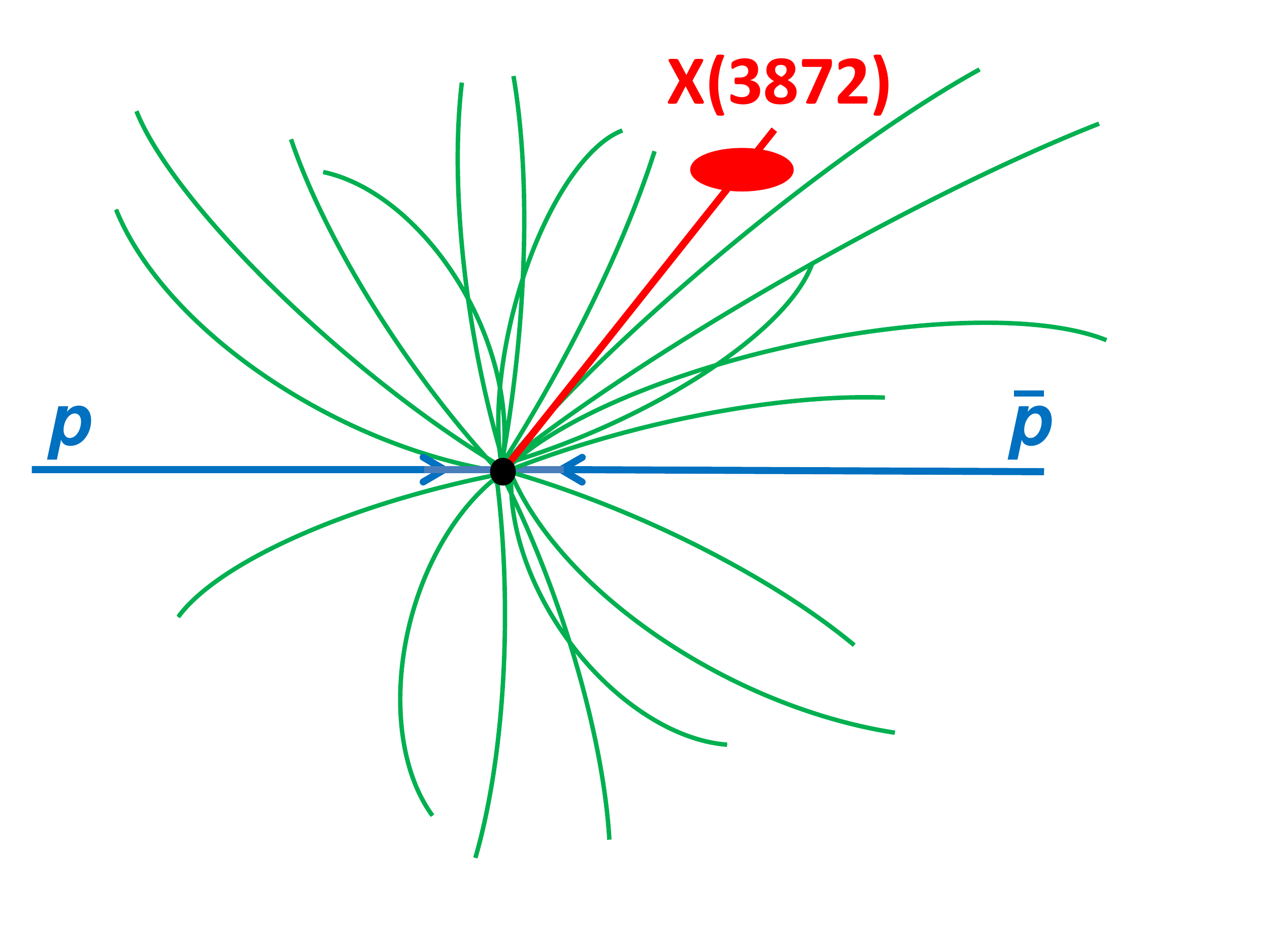}\hspace{1.cm}\includegraphics[width=.35\textwidth]{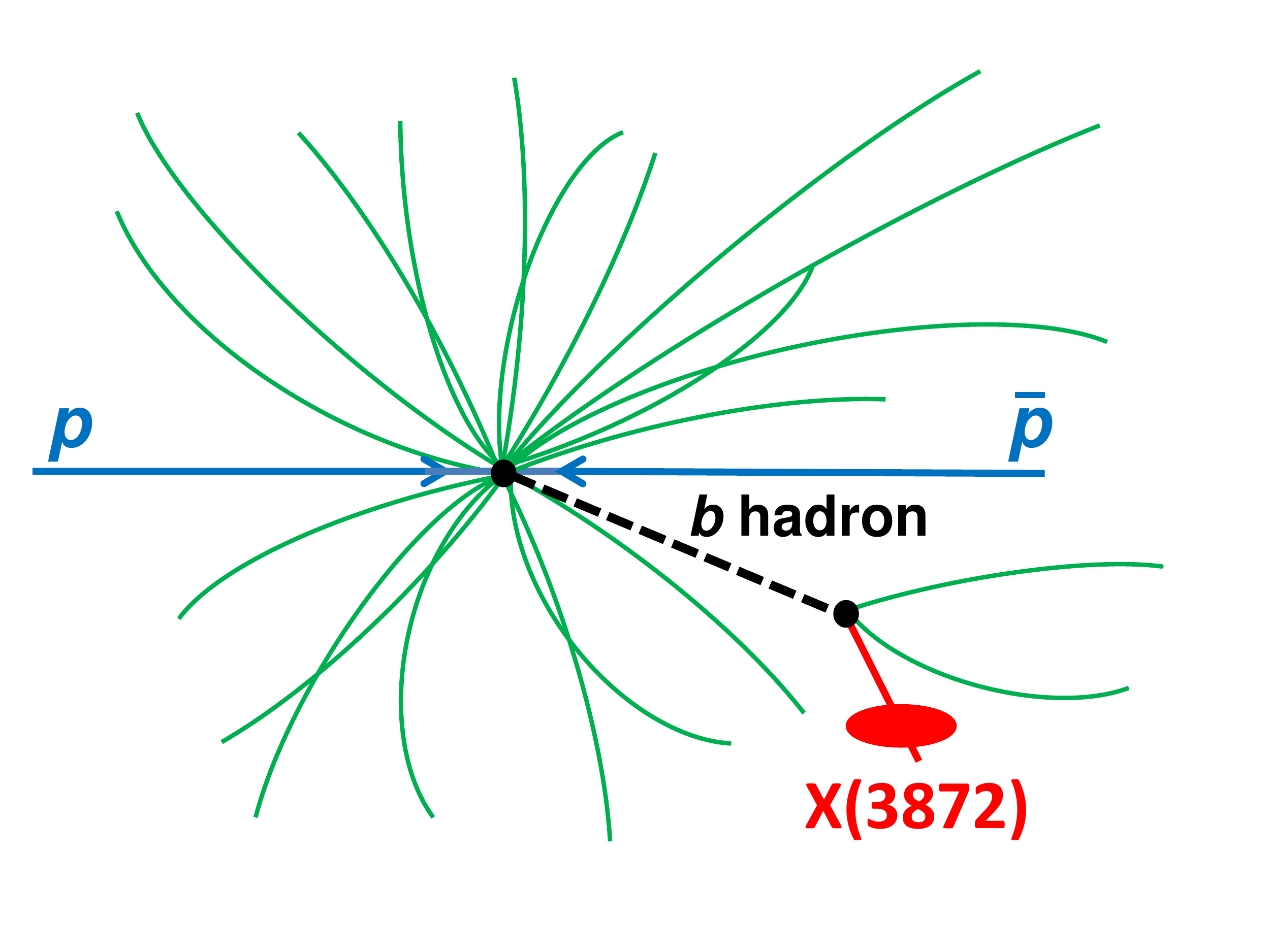}
\end{center}
\vspace{-0.6cm}
\caption{The prompt (left) and nonprompt (right) production of the $X(3872)$ state.}
\label{fig1}
\end{figure}

Many experimental measurements of new exotic states performed during
the last decade stimulated a wide theoretical discussion about 
possible interpretations of these states.
Most of these efforts have been focused on the measurements and
theoretical interpretations of the decays of exotic states,
whereas less attention has been paid to the production mechanism studies. 
At hadron colliders exotic states can be produced nonpromptly in the secondary
vertices of $b$ hadron decays or promptly in hadron 
collisions with many particles coming from the primary vertex (Fig.~\ref{fig1}).

A theoretical discussion of potential production mechanisms was motivated by 
the experimental observation of the copious prompt production
of the $X(3872)$  in hadron colliders.
Prompt production of a loosely bound molecular state like $X(3872)$ in
violent strong interactions is difficult to explain.
These studies indicated  the possibility for getting more insight about 
configurations of exotic states using information about their production
in different processes, in particular in the prompt hadron-hadron collisions.
The large difference in the spatial sizes of conventional, compact multiquark
and molecular states implies suppressed production rates for spatially
extended configurations in the primary vertices at hadron-hadron collisions.
The dependence of the prompt $X(3872)$ production rate on the number
of particles produced in the primary vertex at hadron colliders was
observed by the LHCb experiment~\cite{lhcb} and was theoretically 
discussed in~\cite{diss}, in which the disassociation of spatially extended and
weakly bound $X(3872)$ state by comoving particles was calculated. 
However the production dynamics of a spatially large object in the primary
vertex of hadron-hadron collisions requires additional theoretical investigations.

Up to now, at LHC the only exotic states observed in prompt hadron-hadron collisions
are the $X(3872)$ and the $X(6900)$ state observed recently by the LHCb experiment
in the invariant mass of two $J/\psi$ mesons.
The $X(3872)$ state is often assumed to be a mixture of conventional 
$\chi_{c1}(2P)$ and molecular configurations, therefore the copious 
prompt production could possibly come through its conventional $\chi_{c1}(2P)$ component.
The $X(6900)$ state decaying in two $J/\psi$ mesons consists of four
heavy quarks and should have a small spatial size, so its production 
would not be suppressed. At Tevatron, additionally to the
observation of the prompt $X(3872)$
production, an evidence of 4.7$\,\sigma$ 
was found for the prompt $X(4140)$ signal~\cite{do1} and the promptly
produced $X(5568)$ state 
was observed in the $B_s^0 \pi^{\pm}$ system~\cite{do2}. The $X(5568)$ signal was not
confirmed by LHC experiments.   However the direct comparison of the prompt production
ratios of the spatially extended four-quark $X(5568)$ state to the conventional
$B_s^0$ meson at LHC and Tevatron conditions is not appropriate.
This ratio can be strongly suppressed at LHC compared to the Tevatron measurement,
where about half as many particles on average are produced in the primary 
hadron collision vertex as at LHC.

\section{Prompt and nonprompt production of {\boldmath $X(3872)$} and {\boldmath $\psi(2S)$} states}

The pseudo-proper time distributions for the $X(3872)$ and $\psi(2S)$
states were studied by the D0 collaboration
with a $\sim$10.4~fb$^{-1}$ data sample~\cite{do3}.
The pseudo-proper time $t_{pp}$ is calculated using the formula
$t_{pp} =  \vec L_{\rm xy}\hspace{-0.07cm}\cdot\hspace{-0.05cm}\vec{p}_T \, m  / (p_T^2 \,c)$,
where $\vec{p}_T$ and $m$ are the transverse momentum and mass
of the charmonium state $\psi(2S)$ or $X(3872)$ and $c$ is the speed of light.
To obtain the $t_{pp}$ distributions, the numbers of events are extracted from fits
for the $X(3872)$ ($\psi(2S)$) mass in 12 (24) exponentially increasing
$t_{pp}$ bins.
The  $t_{pp}$ distributions are fitted using
the $\chi^2$ method with a model that includes prompt and nonprompt components.
The prompt production is assumed to have a strictly zero lifetime,
whereas the nonprompt component
is assumed to be distributed exponentially starting from zero.  Both shapes are
smeared by the detector vertex resolution.

\begin{figure}[h]
\begin{center}
\includegraphics[width=.51\textwidth]{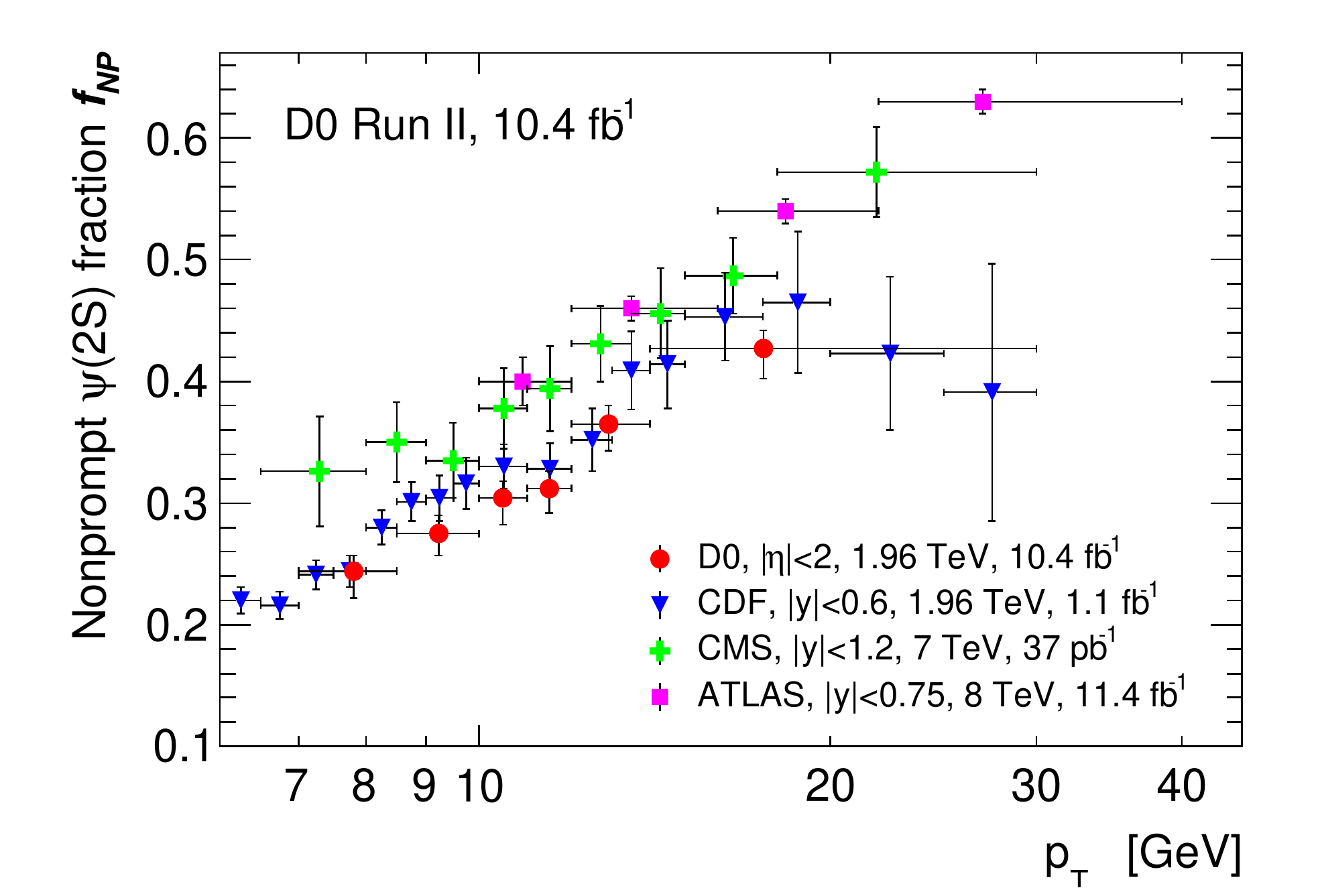}\includegraphics[width=.51\textwidth]{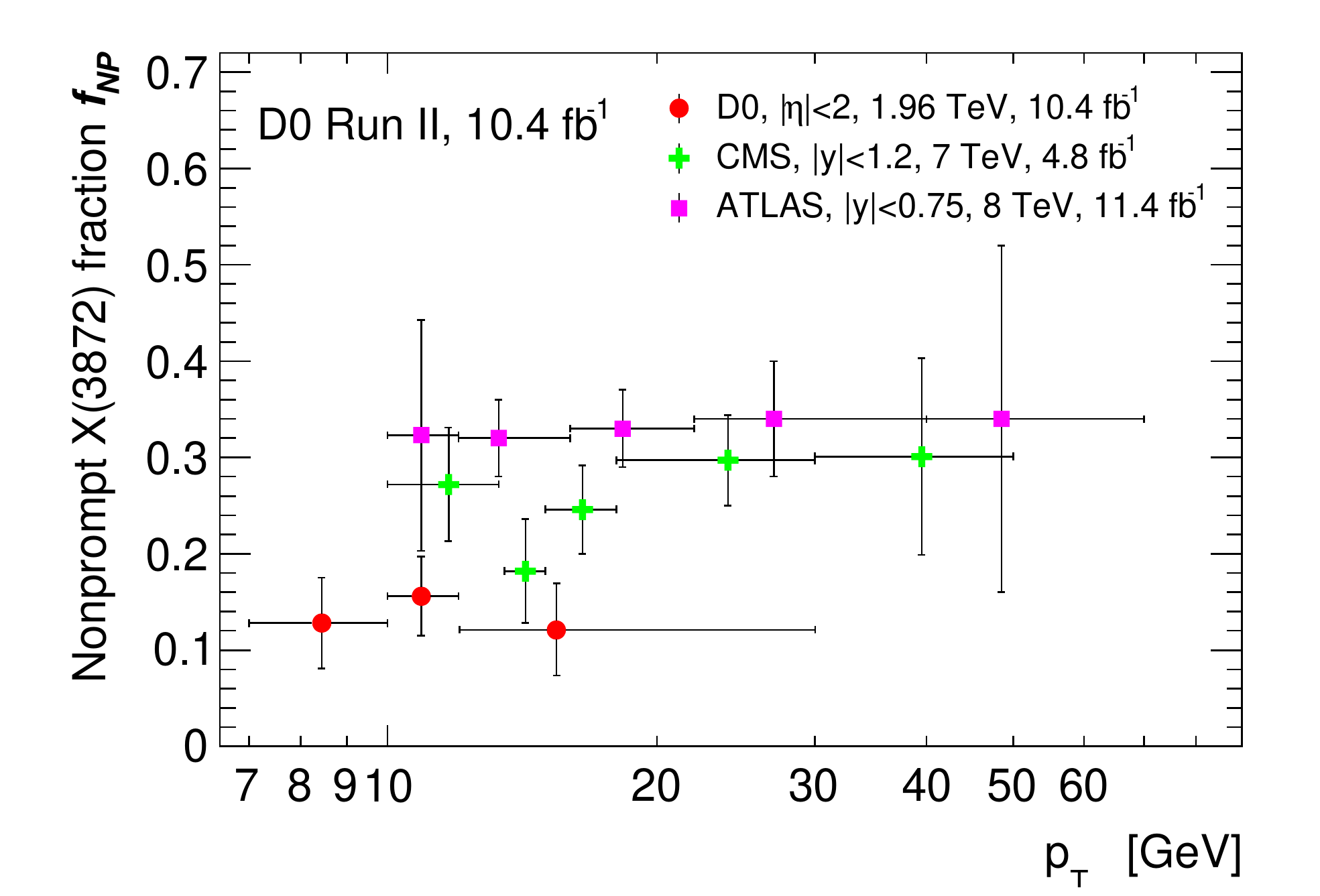}
\end{center}
\vspace{-0.2cm}
\caption{The nonprompt component $f_{NP}$ for the $\psi(2S)$ (left) and
$X(3872)$ (right) states as a function of $p_T$, in comparison with the ATLAS, CMS
and CDF measurements.}
\label{fig2}
\end{figure}

The large D0 sample allows the  study of the $t_{pp}$ distributions
in several $p_T$ intervals and 
the nonprompt contribution fraction $f_{NP}$ can be extracted in each.   Figure~\ref{fig2} (left)
shows the D0 $f_{NP}$ as a function of $p_T$ for the $\psi(2S)$, compared with the measurements by
ATLAS~\cite{atlas} at 8~TeV, CMS~\cite{cms1} at 7~TeV,
and  CDF~\cite{cdf} at 1.96~TeV.
Figure~\ref{fig2} (right) shows similar distributions for the $X(3872)$ obtained
in the D0 analysis, together with the
ATLAS~\cite{atlas} and CMS~\cite{cms2} measurements.

The nonprompt fractions $f_{NP}$ for $\psi(2S)$ increase as a function of $p_T$,
whereas those for $X(3872)$ are consistent with being independent of $p_T$,
similar to the measurements of other experiments.
The ratio of prompt to nonprompt $\psi(2S)$ production,
$R_{p/np} = (1 - f_{NP}) / f_{NP}$,
is only about 25 \% larger at the Tevatron than at the LHC, but for the
$X(3872)$ production at the Tevatron exceeds that at the LHC by about 3.
This indicates that the prompt production of the exotic state $X(3872)$ 
is strongly suppressed at the LHC relative to that from $b$ hadron decays.
This suppression is possibly due to the larger number of particles produced in the primary
collision at the LHC than at the Tevatron, thus increasing the probability for 
disassociating the nearly
unbound and spatially extended $X(3872)$.

Recent theoretical works predict a sizable contribution to the production
of the $X(3872)$ from the formation of the $X(3872)$ in association with
a comoving pion, both directly in the hadronic beam collisions~\cite{brap}
and in $b$ hadron decays~\cite{brab}.
The $X(3872)$ state can be produced by the creation of $D^*\bar D^*$
at short distances, followed by a rescattering of the charm-meson pair
into a $X(3872)\pi$ pair by exchanging a $D$ meson, where the $X\pi$ 
kinetic energy in the $X \pi$ center of mass frame is expected to be $T<11.8$~MeV.

\begin{figure}[h]
\begin{center}
\includegraphics[width=.48\textwidth]{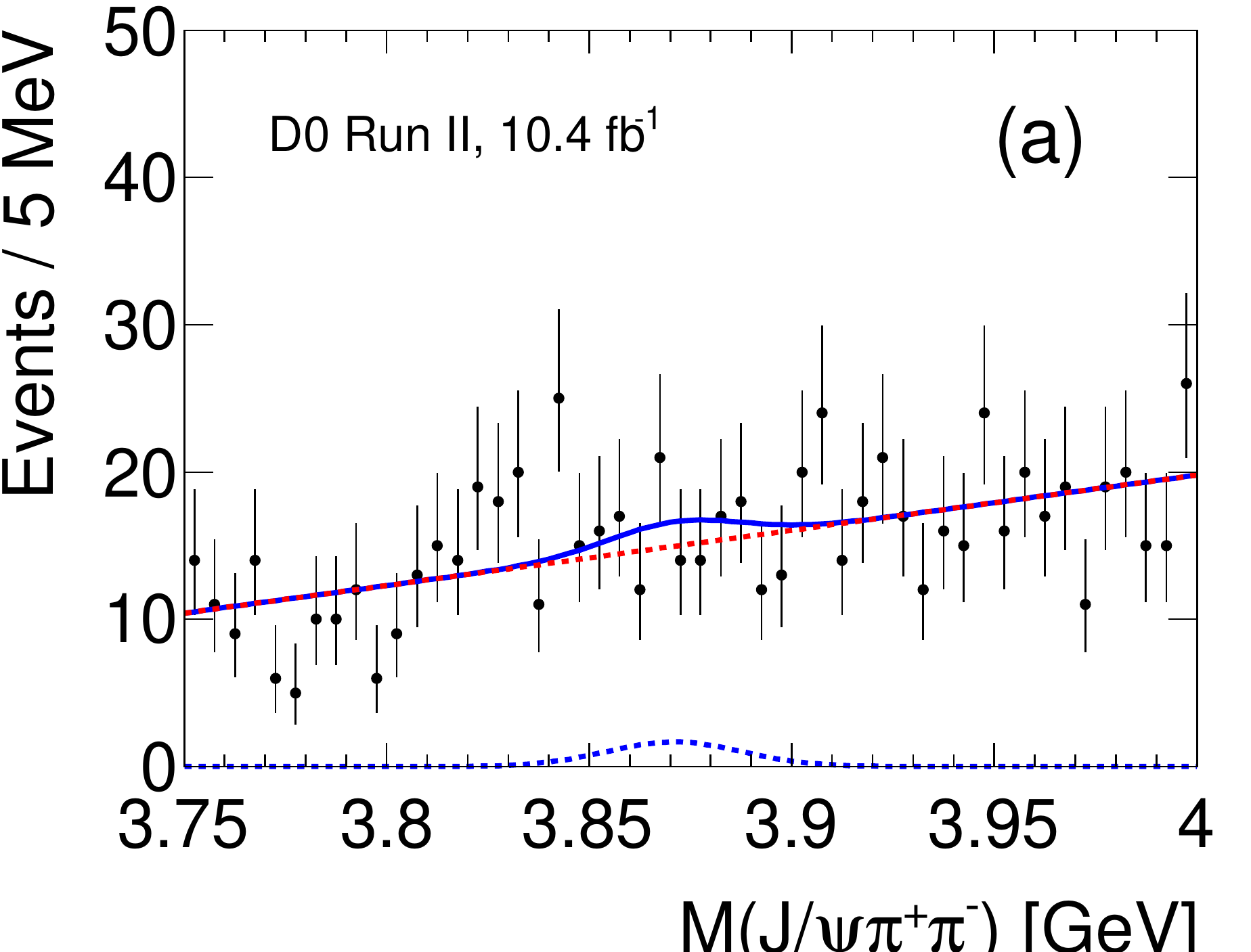}\hspace{0.2cm}\includegraphics[width=.47\textwidth]{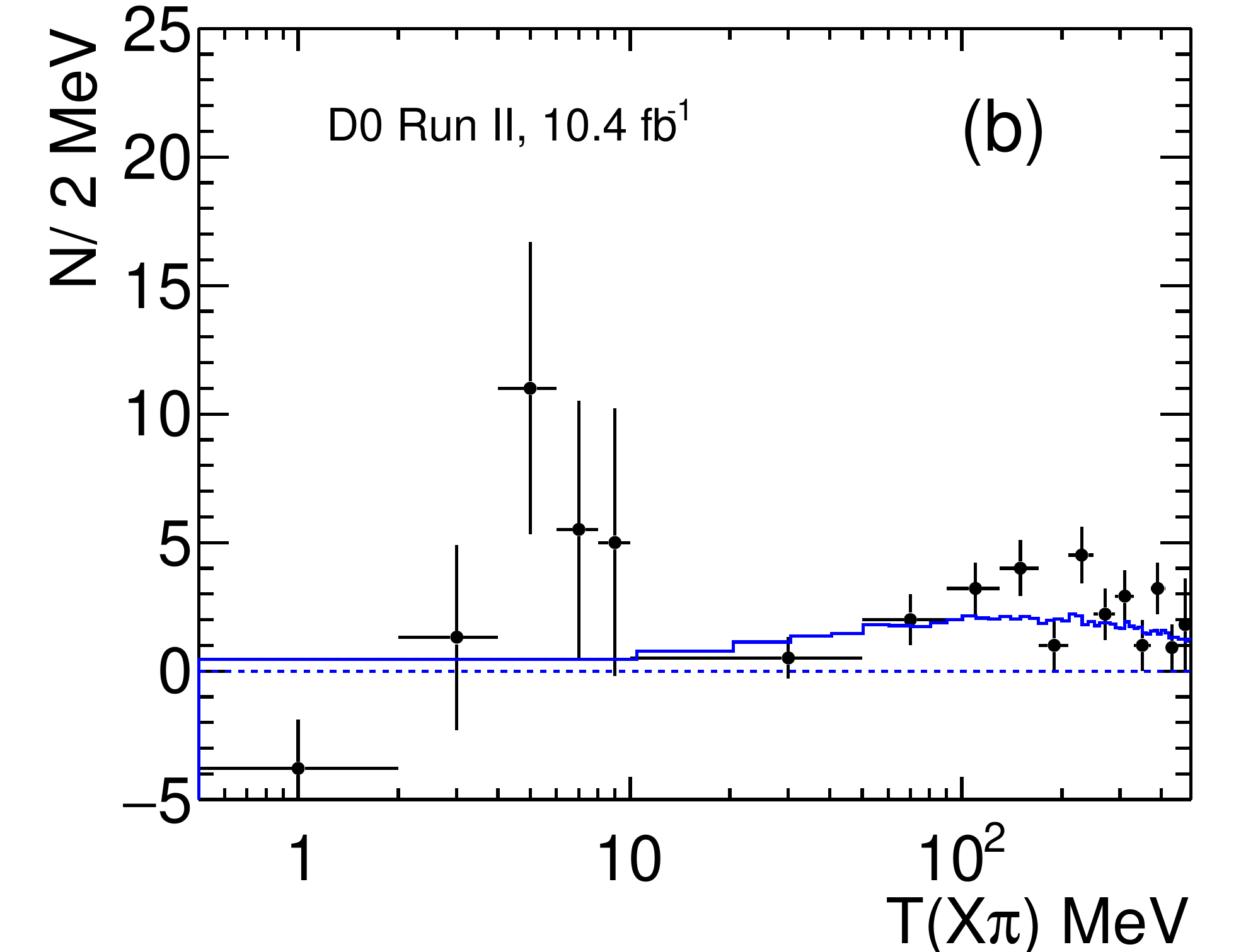}
\end{center}
\caption{(a) The $X(3872)$ signal at the $T(X\pi)<11.8$~MeV region
for prompt events and (b) the $X(3872)$ signal yield as a function of $T(X\pi)$
for nonprompt events.}
\label{fig3}
\end{figure}

The $X(3872)$ signal obtained by D0 using the prompt sample~\cite{do3} within
the $T(X\pi)<11.8$~MeV region is shown in Fig.\ref{fig3} (a).
The fit yields 18$\,\pm\,$16 events. In the absence of the soft-pion process,
6 events are expected in this region, to be compared with an estimated 245--730
events from the soft-pion process.
Therefore no evidence for the soft-pion effect is seen in the prompt sample.
For the nonprompt $X(3872)$ sample, a signal of 27$\,\pm\,$12 events
is observed in the low $T(X\pi)$ region (Fig.\ref{fig3} (b)),
whereas only 2 events are expected from the energy distribution extrapolation.
The expected number of soft-pion events is between 30 and 90.
The observed number of events differs from that without the soft-pion process
only by $2\sigma$, preventing a definite conclusion.

\vspace{-0.3cm}
\section{Prompt and nonprompt production of {\boldmath $Z_c^\pm(3900)$}}

\begin{figure}[h]
\begin{center}
\includegraphics[width=.44\textwidth]{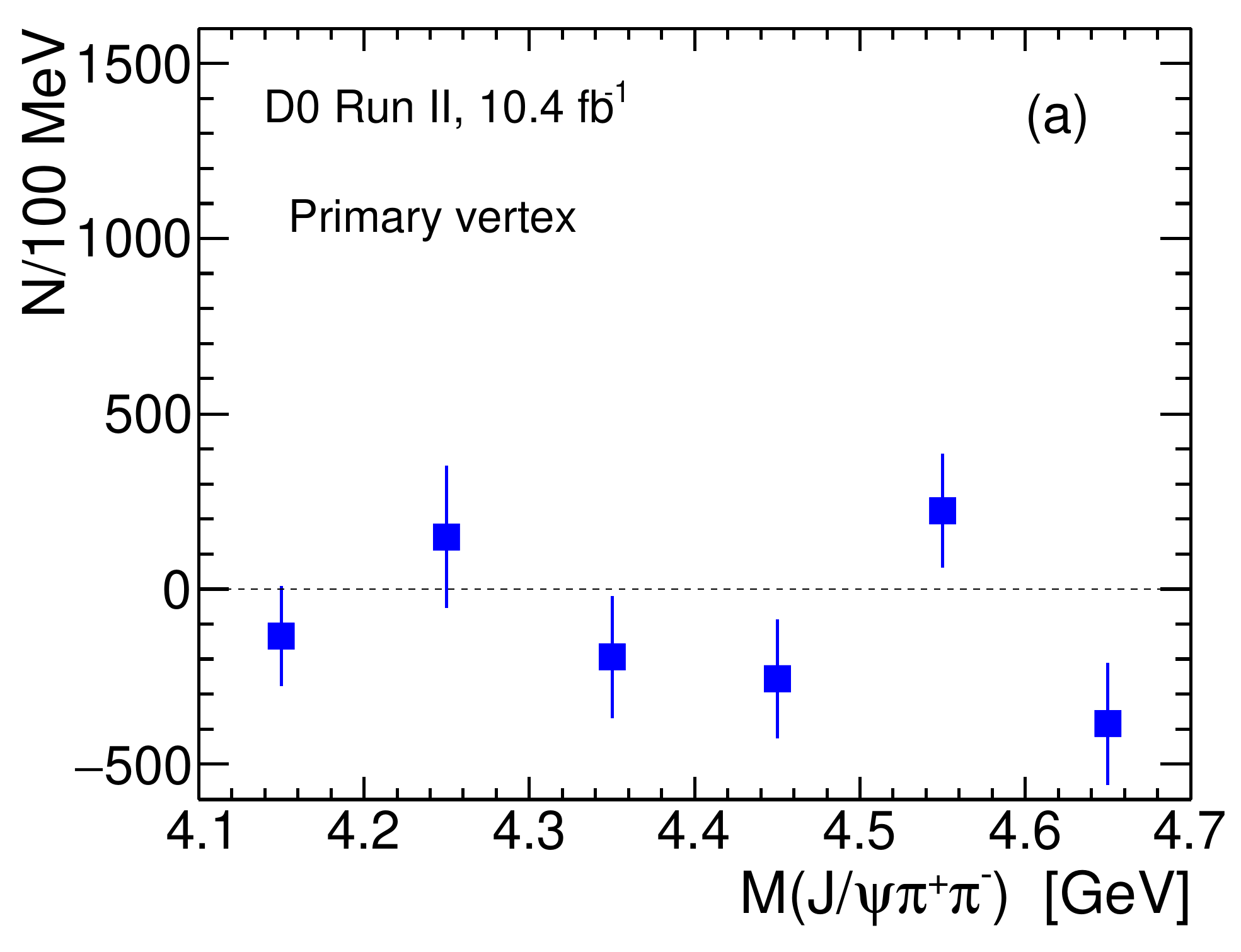}\hspace{0.5cm}\includegraphics[width=.44\textwidth]{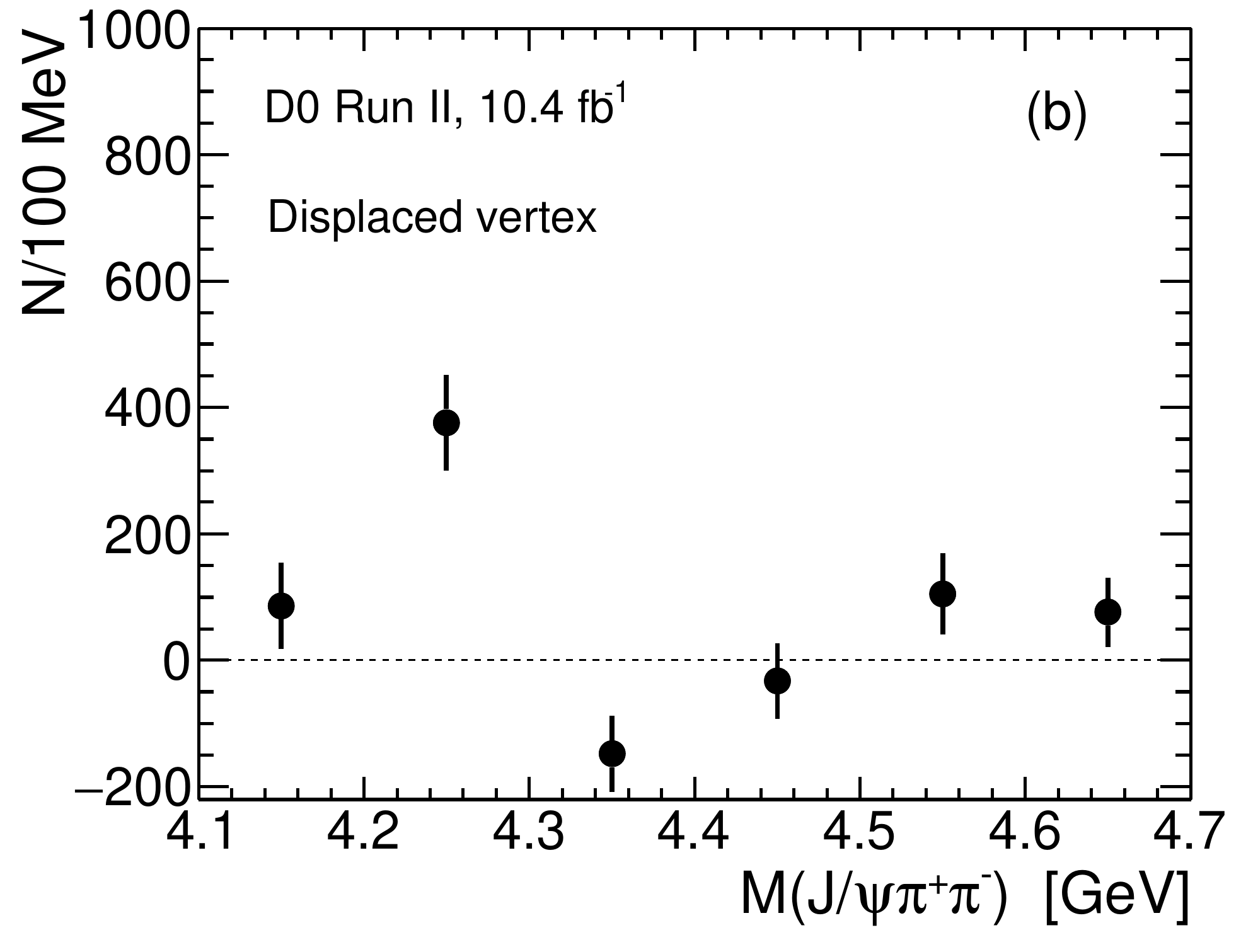}
\end{center}
\vspace{-0.4cm}
\caption{The $Z_c^\pm(3900)$ signal yield obtained from fits for six 1~GeV
wide intervals in the $4.1 < M\,(J/\psi \pi^+\pi^-) < 4.7$ GeV range
for (a) prompt and (b) nonprompt data samples.}
\label{fig4}
\end{figure}

The prompt and nonprompt production of the exotic charged
state $Z_c^\pm(3900)$ was studied by the D0 collaboration~\cite{do4}
through the sequential process 
$\psi(4260) \to Z_c^\pm(3900) \pi^\mp, Z_c^\pm(3900) \to J/\psi \pi^\pm$.
The nonprompt events are selected semi-inclusively, requiring that all final tracks
form a secondary vertex displaced from the primary $p\bar{p}$ collision vertex.
The events are selected in the $M\,(J/\psi \pi^+\pi^-)$ range 4.1--\,4.7~GeV
that includes the exotic $\psi(4260)$ state.
The fits of the $M\,(J/\psi \pi^\pm)$ distributions are performed in the vicinity of
the $Z_c^\pm(3900)$ for the six 1 GeV wide intervals of $M\,(J/\psi \pi^+\pi^-)$.
The fit results are shown in Fig.~\ref{fig4} for the prompt (a) and nonprompt (b)
samples. No signal is observed for the prompt data sample in any
$M\,(J/\psi \pi^+\pi^-)$ interval.
For the nonprompt sample a clear enhancement is seen for the events in the range 
$4.2 < M\,(J/\psi \pi^+\pi^-) < 4.3$ GeV. For events in this range the
$M(J/\psi \pi^\pm)$ distribution fit is performed and the $Z_c^\pm(3900)$
signal is observed with parameters: $M = 3902.6_{-5.0}^{+5.2}$ GeV, 
$\Gamma = 32^{+28}_{-21}$ GeV, and the statistical significance $S = 5.4\,\sigma$.  

\vspace{-0.3cm}
\section{Evidence for inclusive nonprompt production of {\boldmath $P_c$} states}

\begin{figure}[h]
\begin{center}
\includegraphics[width=.48\textwidth]{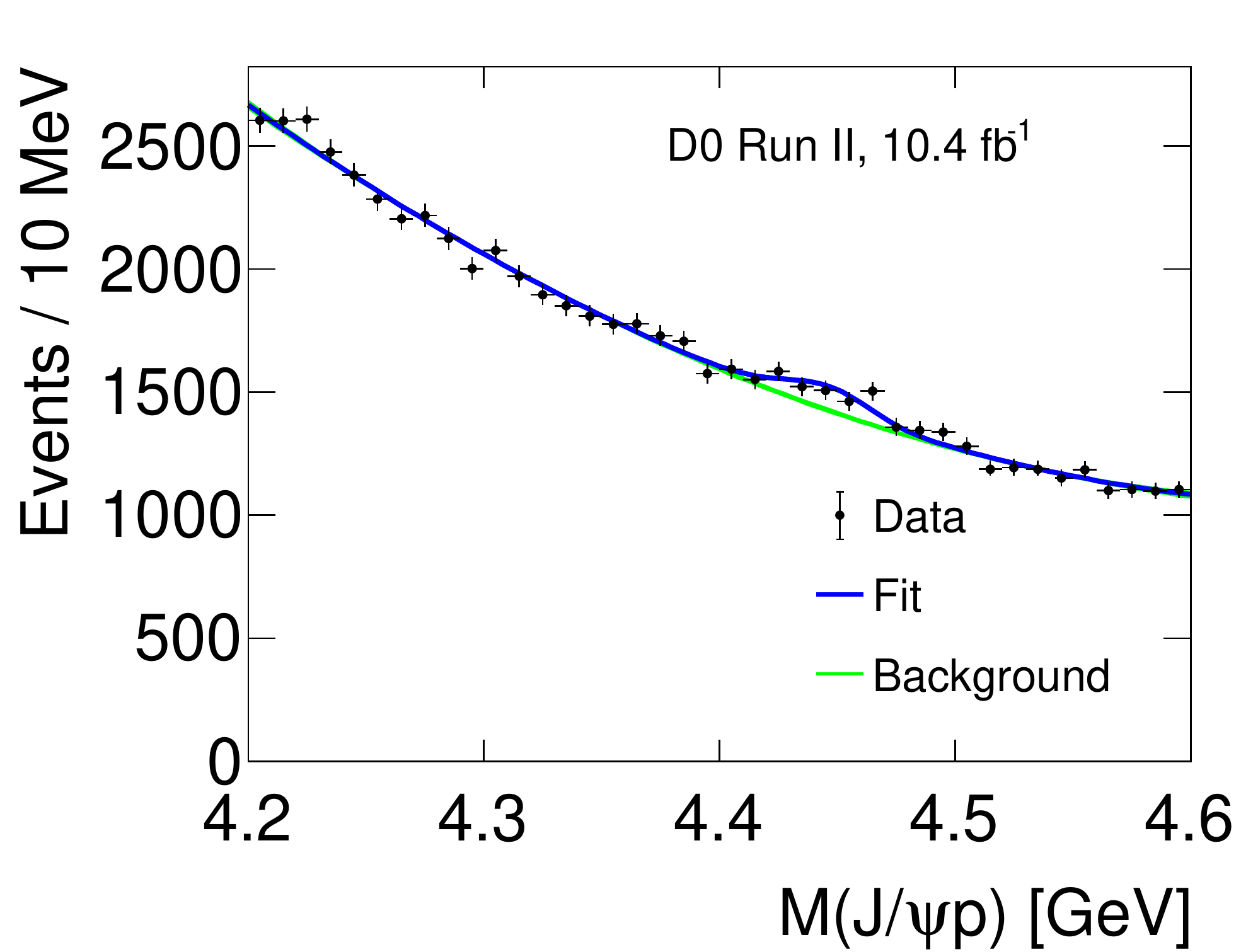}
\end{center}
\vspace{-0.4cm}
\caption{The invariant $M(J/\psi p)$ mass distribution with a superimposed fit,
described in the text.}
\label{fig5}
\end{figure}

The mass spectrum for the $J/\psi p$ combination was studied with the
full D0 data sample~\cite{do5}. 
The preliminary results are obtained for the inclusive production of 
the $J/\psi p$ final state,
where muons from the $J/\psi$ and a proton originate from a common
secondary vertex, displaced in the transverse plane from the
$p\bar{p}$ interaction vertex by at least 5$\,\sigma$.
The invariant $M(J/\psi p)$ mass distribution is shown in Fig.~\ref{fig5}.

In the fit the signal is modeled by a sum of two Breit-Wigner resonances,
corresponding to the $P_c(4400)$ and $P_c(4457)$ pentaquarks, with 
parameters equal to those obtained by the LHCb experiment~\cite{lhcb-pentaq}.
The background is described by a second-order Chebyshev
polynomial. Evidence for the signal is found with a significance 
of 3.2$\,\sigma$.

The author is grateful to Paul Grannis and Daria Zieminska for editing the text.

\end{document}